\documentclass[twocolumn,english,prl,amsmath,amssymb,preprintnumbers]{revtex4}
\usepackage[T1]{fontenc}
\usepackage[latin9]{inputenc}
\usepackage{amstext}
\usepackage{amsmath}
\usepackage{amsfonts}
\usepackage{amssymb}
\usepackage{mathtools}
\usepackage{graphicx}
\usepackage{rotating}
\usepackage {bm}
\usepackage[section]{placeins}
\usepackage{color}
\usepackage{multirow}
\usepackage{cellspace}
\makeatletter
\@ifundefined{textcolor}{}
{%
 \definecolor{BLACK}{gray}{0}
 \definecolor{WHITE}{gray}{1}
 \definecolor{RED}{rgb}{1,0,0}
 \definecolor{GREEN}{rgb}{0,1,0}
 \definecolor{BLUE}{rgb}{0,0,1}
 \definecolor{CYAN}{cmyk}{1,0,0,0}
 \definecolor{MAGENTA}{cmyk}{0,1,0,0}
 \definecolor{YELLOW}{cmyk}{0,0,1,0}
}

\def\Im{{\text{Im}}\,}

\def\kF{k_{\text{F}}}
\def\vF{v_{\text{F}}}

\def\muBSSL VPN Service{\mu_{\text{B}}}
\def\NF{N_{\text{F}}}

\def\epsilonF{\epsilon_{\text{F}}}

\def\sgn{{\text{sgn\,}}}

\def\be{\begin{equation}}
\def\ee{\end{equation}}
\def\bea{\begin{eqnarray}}
\def\eea{\end{eqnarray}}
\def\bse{\begin{subequations}}
\def\ese{\end{subequations}}

\usepackage{dcolumn}
\usepackage{bm}
\usepackage{amsfonts}
\draft

\makeatother

\usepackage{babel}
\begin{document}

\title{Non-Hydrodynamic Initial Conditions are Not Soon Forgotten}

\author{T.R. Kirkpatrick$^{1}$, D. Belitz$^{2,3}$ and J.R. Dorfman$^{1}$}

\affiliation{$^{1}$ Institute for Physical Science and Technology, University of Maryland, College Park, MD 20742, USA\\
                 $^{2}$ Department of Physics and Institute for Fundamental Science, University of Oregon, Eugene, OR 97403, USA\\
                 $^{3}$ Materials Science Institute, University of Oregon, Eugene, OR 97403, USA
                  }

\date{\today}
\begin{abstract}
Solutions to hydrodynamic equations, which are used for a vast variety of physical problems, are assumed to be specified
by boundary conditions and initial conditions on the hydrodynamic variables only. Initial values of other variables are
assumed to be irrelevant for a hydrodynamic description. This assumption is not correct because of the
existence of long-time-tail effects that are ubiquitous in systems governed by hydrodynamic equations.
We illustrate this breakdown of a hydrodynamic description by means of the simple example of
diffusion in a disordered electron system.

\end{abstract}
%
%
\maketitle
Hydrodynamic descriptions of matter have a long history, with classical 
fluids the most obvious example 
\cite{Landau_Lifshitz_VI_1987}. Others include plasmas \cite{Landau_Lifshitz_X_1981, Belmont_et_al_2019},
superfluids, spin transport in magnets, 
and excitations in liquid crystals and in solids \cite{Forster_1975, Martin_Parodi_Pershan_1972}.
More recently, hydrodynamic descriptions have been used to study 
active matter \cite{Marchetti_et_al_2013}, 
general relativity 
\cite{Minwalla_Hubeny_Rangamani_2011}, supersymmetric field
theories and quantum gravity \cite{Kovtun_Yaffe_2003}, quark-gluon 
plasmas \cite{Mazeliauskas_Berges_2019}, 
and the unusual properties of Weyl and Dirac metals and semi-metals \cite{Zaanen_2016}. 

The basic assumption of any hydrodynamic theory is that the behavior of a macroscopic system on length and time
scales large compared to the microscopic ones can be described in terms of a small number of `hydrodynamic'
variables, while all other degrees of freedom are effectively integrated out. 
One of the simplest hydrodynamic processes
is diffusion, with applications in Physics, Chemistry, Biology, and Engineering \cite{Mehrer_2007, Cussler_1997, Stein_1986}. 
It is 
phenomenologically described by
Fick's law, which states that
the current density $\bm j$ associated with 
a conserved density $n$ is proportional to the gradient of $n$,
${\bm j}({\bm x},t) = -D\,{\bm\nabla}n({\bm x},t)$, with $D$ the diffusion coefficient. Together with the continuity 
equation for $n$, this 
leads to the diffusion equation
\be
\partial_t n({\bm x},t) - D\,{\bm\nabla}^2 n({\bm x},t) = 0\ .
\label{eq:1}
\ee
This is expected to be valid for times $t\gg t_0$ and wave numbers $q \ll q_0$,
with $t_0$ and $1/q_0$ the microscopic time and length scales
that depend on the nature of the diffusive process.
In this description there is only one hydrodynamic variable, viz., the density.
With $n_0$ the constant equilibrium density, Eq.~(\ref{eq:1}) describes the relaxation of a
density perturbation $\delta n({\bm x},t) = n({\bm x},t) - n_0$. 
It is solved by a spatial Fourier transform and a temporal Laplace transform \cite{Forster_1975, Laplace_trafo_footnote}. 
The result is the well known diffusion pole in the complex-frequency ($z$) plane,
\bse
\label{eqs:2}
\be
\delta n({\bm q},z) = \delta n({\bm q},t=0)\,i/[z + i D q^2 s_z]
\label{eq:2a}
\ee
with ${\bm q}$ the wave vector, 
 and $s_z = \sgn \Im (z)$. In the time domain, this
corresponds to exponential decay with a relaxation time that diverges as $1/q^2$ 
for small ${\bm q}$,
\be
\delta n({\bm q},t) = \delta n({\bm q},t=0)\,e^{-Dq^2 t}\ .
\label{eq:2b}
\ee
\ese
According to 
Eq.~(\ref{eq:2b}), the initial value 
$\delta n({\bm q},t=0)$ completely
determines the relaxation for $t\gg t_0$ and 
$q \ll q_0$. 

As we will show, the relaxation behavior described by the diffusion
equation is not complete. First, 
kinetic theory shows that
$\delta n({\bm q},t)$ also depends on initial conditions in higher angular-momentum channels; e.g.,
an initial current. 
Within 
diffusion theory, 
these initial conditions are `non-hydrodynamic' since they involve quantities other than the
density. As we will show, solving the standard Boltzmann equation for the scattering 
of electrons by static point-like impurities 
yields
\bse
\label{eqs:3}
\be
\delta n({\bm q},t\gg t_0) = \left[\delta n({\bm q},t=0) + \delta n_{\perp}({\bm q},t=0)\right]\,e^{-Dq^2 t}\ .
\label{eq:3a}
\ee
Here the diffusion coefficient is $D = \vF^2\tau/3$, with $\vF$ the Fermi velocity and
$\tau$ the elastic mean-free time. In this case, the microscopic time and wave number are
$t_0 = \tau$ and $q_0 = 1/\vF\tau$, respectively. 
$\delta n_{\perp}({\bm q},t=0)$ is a set of non-hydrodynamic initial conditions that 
are orthogonal in angular-momentum ($l$) space to the density, with the 
current density ${\bm j}$ ($l = 1$) the leading contribution,
\be
\delta n_{\perp}({\bm q},t=0) \approx -i \tau{\bm q}\cdot{\bm j}({\bm q},t=0)\ .
\label{eq:3b}
\ee
The density relaxation is still given by the solution of the diffusion equation,
Eq.~(\ref{eq:2b}), if we 
replace $\delta n(t=0)$ by $\delta n$ taken
at an adjusted initial time $t_s$ defined by
\bea
\delta n({\bm q},t_s) &\equiv& \delta n({\bm q},t=0)\,e^{-Dq^2 t_s} 
\nonumber\\
&=& \delta n({\bm q},t=0) + \delta n_{\perp}({\bm q},t=0)\ .
\label{eq:3c}
\eea
\ese
The  `slip time' $t_s$ is the temporal analog of the `slip length' used to describe
the flow of a fluid near a surface, i.e., the distance between the actual surface and the
fictitious one where hydrodynamic boundary conditions can be 
used \cite{Dorfman_vanBeijeren_Kirkpatrick_2021, Nieuwoudt_Kirkpatrick_Dorfman_1984}.
If the initial conditions are such that $\delta n$ and $\delta n_{\perp}$ are related
by Fick's law, $\delta n_{\perp}({\bm q},t=0) \approx D q^2\tau \delta n({\bm q},t=0)$,
then $\vert t_s\vert \approx \tau$, but generically this will not be the case.

Second, the exponential decay predicted by 
Eq.~(\ref{eq:3a}) is not the true
asymptotic long-time behavior:  Correlation 
effects lead to a non-analytic frequency dependence
of the diffusion coefficient. 
As a function of time, this corresponds to a power-law decay of 
$\delta n$. 
Such non-exponential `long-time tails' (LTTs) are not described by the Boltzmann equation 
\cite{Dorfman_Cohen_1970, Ernst_Hauge_van_Leeuwen_1970},
yet are present in all systems described by hydrodynamics 
\cite{Kovtun_Yaffe_2003, Marchetti_et_al_2013, Dorfman_vanBeijeren_Kirkpatrick_2021}.

In this Letter we show that the combination of LTTs and
non-hydrodynamic initial conditions yields a long-time behavior of the density relaxation
that cannot be reconciled with the diffusion equation by adjusting the initial condition as in 
Eq.~(\ref{eq:3c}), even 
with a 
frequency-dependent diffusivity that yields the correct LTT. 

As an example, we consider the density relaxation of noninteracting conduction electrons in a three-dimensional 
(3-d) system 
with weak quenched disorder. The latter leads to LTTs
known as weak-localization (WL) effects \cite{Lee_Ramakrishnan_1985, AA_effects_footnote}, but in 3-d systems
it does not lead to Anderson localization; the system remains metallic. Let $\epsilonF$ and $\kF$ 
be the Fermi energy and wave number, 
respectively, and $m$ the effective mass. 
The density of states per spin at the Fermi surface is
$\NF = \kF m/2\pi^2$, and the mean-free path is $\ell = \vF\tau$. For weak disorder ($\kF\ell \gg 1$, or $\epsilonF\tau \gg 1$), and 
and $t \gg 1/Dq^2$, we find 
\bse
\label{eqs:4}
\bea
\delta n({\bm q},t) &=& \delta n({\bm q},t=0)\left[\frac{3}{4\sqrt{\pi}}\,\frac{1}{\kF\ell}\,\frac{q/\kF}{(Dq^2 t)^{3/2}} + O(\frac{1}{t^{5/2}})\right]
\nonumber\\
&& + \delta n_{\perp}({\bm q},t=0)\,\frac{9}{8\sqrt{\pi}}\,\frac{1}{\kF\ell}\,\frac{q/\kF}{(Dq^2 t)^{5/2}}\ .\qquad
\label{eq:4a}
\eea
The different LTTs in the two terms make it
impossible to account for the non-hydrodynamic initial conditions represented by $\delta n_{\perp}(t=0)$
by means of a slip time as in Eqs.~(\ref{eqs:3}). A diffusion pole, Eq.~(\ref{eq:2a}), 
generalized to allow for the LTT, thus fails to describe the density relaxation at long times. 
This remains true for intermediate times $\tau \ll t \ll 1/Dq^2$, where the leading time dependence
(ignoring an $O(1/\kF\ell)$ contribution to the constant part of the $\delta n_{\perp}$ term) is
\bea
\delta n({\bm q},t) &=& \delta n({\bm q},t=0)\,\left[1 + O(D q^2 t)\right]
\nonumber\\
&&\hskip -50pt + \delta n_{\perp}({\bm q},t=0)\,\left[1 - \frac{3}{2\sqrt{\pi}}\,\frac{1}{\kF\ell}\,\frac{q/\kF}{(Dq^2 t)^{1/2}}\right]\ .
\label{eq:4b}
\eea
\ese
Here the difference between the hydrodynamic and non-hydrodynamic terms is even more striking
than in Eq.~(\ref{eq:4a}): The 
former is a constant, 
as described by the diffusion equation, whereas the 
latter
has a $1/t^{1/2}$ long-time tail. Again, the concept of a slip time breaks down.

Equations~(\ref{eqs:4}), and the related comments, represent our main result which we will now
derive from kinetic theory. The dependence on non-hydrodynamic
initial conditions arises already at the level of the Boltzmann equation, while the WL
LTTs require a more sophisticated treatment of collision processes.

\medskip
{\it Kinetic equation for the single-particle distribution}\ 
Consider the single-electron distribution function $f({\bm p},{\bm x},t)$ as a function of the
electron momentum ${\bm p}$, real-space position ${\bm x}$, and time $t$. In the absence
of external forces, the kinetic equation that governs 
$f$ reads
\be
\left(\partial_t + {\bm p}\cdot\bm\nabla_{\bm x}/m\right) f({\bm p},{\bm x},t) =  \left(\partial f/\partial t\right)_{\text{coll}}\ .
\label{eq:5}
\ee
This is completely general:
The total time derivative of $f$ on the left-hand side equals the collision
integral, i.e., the temporal change of $f$ due to collision, on the right-hand side. The equilibrium
distribution function is given by the Fermi-Dirac distribution $f_0({\bm p}) = 1/(e^{(\epsilon_{\bm p} - \mu)/T)}+1)$,
with $\epsilon_{\bm p} = {\bm p}^2/2m$ the single-electron energy, $\mu$ the chemical potential, and $T$ the temperature. 
We parameterize the deviation from equilibrium, $\delta f = f - f_0$, 
by
\be
\delta f({\bm p},{\bm x},t) = w(\epsilon_{\bm p})\,\phi({\bm p},{\bm x},t)\ ,
\label{eq:6}
\ee
where
$w(\epsilon_{\bm p}) = -\partial f_0/\partial\epsilon_{\bm p}$ 
is a weight function. 
For later reference, the number density $n$ and the number current density ${\bm j}$ are
\be
n({\bm x},t) = \frac{1}{V}\sum_{\bm p} f({\bm p},{\bm x},t)\ ,\ 
{\bm j}({\bm x},t) = \frac{1}{V}\sum_{\bm p} {\bm p}\, f({\bm p},{\bm x},t).
\label{eq:6.1}
\ee

\medskip
{\it Boltzmann equation\ } The Boltzmann collision integral for elastic scattering by impurities is \cite{Kohn_Luttinger_1957, Greenwood_1958}
\bse
\label{eqs:7}
\bea
\left(\frac{\partial f}{\partial t}\right)_{\text{coll}}^{\text{B}} &=&  \frac{-1}{\NF V}\sum_{{\bm p}'} \delta(\epsilon_{\bm p} - \epsilon_{{\bm p}'})
     \,\frac{1}{\tau}\,W(\hat{\bm p},\hat{\bm p}')
\nonumber\\
&& \times 
\left[\delta f({\bm p},{\bm x},t) - \delta f({\bm p}',{\bm x},t)\right].
\label{eq:7b}
\eea
The delta-function reflects the elastic nature of the collisions, 
and $W$ is a form factor. For point-like scatterers, the latter is a constant equal to one and we
obtain 
\be
\left(\frac{\partial f}{\partial t}\right)_{\text{coll}}^{\text{B}} = \frac{-1}{\tau}\left[\delta f({\bm p},{\bm x},t) - \delta{\bar f}(\epsilon_{\bm p},{\bm x},t)\right]\ ,
\label{eq:7c}
\ee
where $\delta{\bar f}(\epsilon_{\bm p},{\bm x},t) = \sum_{{\bm p}'} \delta(\epsilon_{\bm p} - \epsilon_{{\bm p}'})\,\delta f(\epsilon_{\bm p},{\bm x},t)/V\NF$ is $\delta f$ averaged over the $\epsilon_{\bm p}$-energy shell.
Integrating over the momentum
yields $\sum_{\bm p} \delta{\bar f}(\epsilon_{\bm p},{\bm x},t) =\sum_{\bm p} \delta f({\bm p},{\bm x},t)$, 
so the 
number density, Eq.~(\ref{eq:6.1}), 
is conserved. 
\ese
After a Fourier-Laplace transform as in Eq.~(\ref{eq:2a}),
the linearized Boltzmann equation 
reads 
\bea
\left(-i z + \frac{i}{m}\,{\bm p}\cdot{\bm q} + \frac{1}{\tau}\right) \phi({\bm p},{\bm q},z) &=& \phi({\bm p},{\bm q},t=0)
\nonumber\\
&& +  \frac{1}{\tau}\,\bar\phi(\epsilon_{\bm p},{\bm q},z)\ , \quad\ 
\label{eq:8}
\eea
with $\phi$ from Eq.~(\ref{eq:6}) and $\bar\phi$ defined in analogy to $\delta{\bar f}$.
Now consider $\delta n = n - n_0$, the deviation of the particle-number
density from its equilibrium value $n_0$. 
From Eq.~(\ref{eq:8}) we find
\bse
\label{eqs:9}
\be
\delta n({\bm q},z) = J_0({\bm q},z) + \frac{i}{\tau}\,\frac{1}{V}\sum_{\bm p} \frac{w(\epsilon_{\bm p}){\bar\phi}(\epsilon_{\bm p},{\bm q},z)}{z - {\bm p}\cdot{\bm q}/m + i/\tau}\ ,
\label{eq:9a}
\ee
where 
\be
J_0({\bm q},z) = \frac{i}{V} \sum_{\bm p} \frac{\delta f({\bm p},{\bm q},t=0)}{z - {\bm p}\cdot{\bm q}/m + i/\tau}\ .
\label{eq:9b}
\ee
\ese
For $T\ll\epsilonF$ we can replace ${\bm p}\cdot{\bm q}/m$ in Eq.~(\ref{eq:9a}) by $\vF{\hat{\bm p}}\cdot{\bm q}$, 
which neglects corrections of $O(T^2)$ to the 
${\bm q}$-dependence of $\delta n$. The resulting closed equation for $\delta n$ yields
%
\bse
\label{eqs:10}
\be
\delta n({\bm q},z) = J_0({\bm q},z)/(1 - iJ({\bm q},z)/\tau)\ ,
\label{eq:10a}
\ee
where
\be
J({\bm q},z) = \frac{1}{4\pi} \int d\Omega_{\bm p}\ \frac{1}{z - \vF{\hat{\bm p}}\cdot{\bm q} + i/\tau}\ ,
\label{eq:10b}
\ee
\ese
with $\Omega_{\bm p}$ the solid angle associated with ${\bm p}$. 
An expansion in the limit of small wave numbers ($q\ell \ll 1$) and frequencies ($z\tau \ll 1$) 
yields
\bse
\label{eqs:11}
\bea
J_0({\bm q},z) = \frac{i\delta n({\bm q},t=0)}{z + i/\tau}\, + \frac{i{\bm q}\!\cdot\!{\bm j}({\bm q},t=0)}{(z + i/\tau)^2}\, + O(q^2),\quad
\label{eq:11a}\\
1 - \frac{i}{\tau}\,J({\bm q},z) = -i\tau(z + i D_0 q^2) + O(z^2,q^4)\ , \qquad
\label{eq:11b}
\eea
\ese
where $D_0 = \vF^2\tau/3$ is the Boltzmann diffusion coefficient. Inserting these results into
Eq.~(\ref{eq:10a}), and performing a Laplace back transform, yields Eqs.~(\ref{eq:3a}, \ref{eq:3b}).

Equations~(\ref{eq:9b}) and (\ref{eq:11a}) show that the solution depends, in addition to 
the initial density perturbation, on the initial current density, as well as higher modes. At the level
of the Boltzmann equation, after a few mean-free times the various initial-condition terms all multiply the same time dependence
and therefore can be incorporated into a `slip time' as in Eq.~(\ref{eq:3c}). This is no longer the 
case if one uses a more sophisticated collision integral that accounts for LTTs. 

\medskip
{\it Weak-localization effects} The LTTs associated with WL effects 
arise from two-particle correlations that are not included in the Boltzmann collision integral
and lead to a frequency-dependent diffusivity that is nonanalytic at
zero frequency \cite{Abrahams_et_al_1979, Wegner_1979}. Reference \onlinecite{Hershfield_Ambegaokar_1986}
showed that these effects can be incorporated into the kinetic equation for the single-particle
distribution by adding an additional term to the 
collision integral. This 
additional %
term 
can still be written in the form of Eq.~(\ref{eq:7b}) 
\cite{Strinati_Castellani_DiCastro_1989},
with two modifications: First, the scattering rate $1/\tau$ must be replaced by a
time-dependent memory function $\alpha$ (to be specified below) that encodes correlations. The WL contribution to the
collision integral 
then has the form
\bea
\left(\frac{\partial f}{\partial t}\right)_{\text{coll}}^{\text{WL}} &=& - \int_{0}^{t} dt' \alpha(t-t')\, \frac{1}{\NF V}\sum_{{\bm p}'} \delta(\epsilon_{\bm p} - \epsilon_{{\bm p}'})
\nonumber\\
&& \hskip -30pt \times W^{\text{WL}}(\hat{\bm p},\hat{\bm p}')\, \left[\delta f({\bm p},{\bm x},t') - \delta f({\bm p}',{\bm x},t')\right]\ .
\label{eq:12}
\eea
Second, the form factor is strongly angle dependent, which reflects the fact that the WL effects
are due to backscattering events \cite{Gorkov_Larkin_Khmelnitskii_1979}:
\bse
\label{eqs:13}
\be
W^{\text{WL}}(\hat{\bm p},\hat{\bm p}') = \frac{1}{4\pi}\,\delta(\hat{\bm p} + \hat{\bm p}') - 1\ ,
\label{eq:13a}
\ee
The additional contribution to the collision integral does not change the total scattering rate \cite{Strinati_Castellani_DiCastro_1989}:
\be
\int d\Omega_{{\bm p}'}\, W^{\text{WL}}(\hat{\bm p},\hat{\bm p}') = 0\ .
\label{eq:13b}
\ee
\ese
The WL contribution to the collision integral becomes
\bse
\label{eqs:14}
\be
\left(\frac{\partial f}{\partial t}\right)_{\text{coll}}^{\text{WL}} = \int_{0}^{t} dt' \alpha(t-t') \left[\delta f({-\bm p},{\bm x},t') - \delta{\bar f}(\epsilon_{\bm p},{\bm x},t')\right].
\label{eq:14a}
\ee
The function $\alpha(t)$ 
and its Laplace transform are \cite{alpha_footnote, mode-coupling_footnote}
\bea
\alpha(t) &=& \frac{1}{\pi\NF\tau}\,\frac{1}{V}\sum_{\bm k} e^{-D_0 k^2 t} \qquad (t>0)\ .
\label{eq:14b}\\
\alpha(z) &=& \frac{i}{\pi\NF\tau}\,\frac{1}{V} \sum_{\bm k} \frac{1}{z + i D_0 k^2 s_z}
\label{eq:14c}\\
  &=& s_z\,\frac{1}{\tau}\,\frac{3}{\pi}\,\frac{1}{\kF\ell}\left[c - \frac{\pi}{2}\,\zeta^{1/2} + O(\zeta)\right]\ ,
\label{eq:14d}
\eea
\ese
where $\zeta = z\,s_z/i D_0 \kF^2$, $c$ is the non-universal zero-frequency contribution \cite{alpha_footnote},
and Eq.~(\ref{eq:14d}) is valid for $d=3$.
Note the 
nonanalytic frequency dependence that results from integrating over the diffusion pole. Only this contribution to
$\alpha$ matters for the arguments that follow.

The kinetic equation that generalizes Eq.~(\ref{eq:8}) now reads 
\bea
\left(-i z + \frac{i}{m}\,{\bm p}\cdot{\bm q} + \frac{1}{\tau}\right) \phi({\bm p},{\bm q},z) - \alpha(z)\,\phi(-{\bm p},{\bm q},z)
&=& 
\nonumber\\
&& \hskip -200pt \phi({\bm p},{\bm q},t=0) + \left(\frac{1}{\tau} - \alpha(z)\right)\,\bar\phi(\epsilon_{\bm p},{\bm q},z) \ . 
\label{eq:15}
\eea
By letting ${\bm p} \to -{\bm p}$ we obtain two coupled equations for $\phi({\bm p})$ and $\phi(-{\bm p})$
that can be solved exactly. 
The result for the density relaxation generalizes Eq.~(\ref{eq:10a}) to
\bse
\label{eqs:16}
\be
\delta n({\bm q}, z) = \frac{{\tilde J}_0({\bm q},z) + i \alpha(z) K_0({\bm q},z)}{1 - [i/\tau -i\alpha(z)] [{\tilde J}({\bm q},z) + i\alpha(z) K({\bm q},z)]}.
\label{eq:16a}
\ee
In terms of a denominator
\be
N({\bm p},{\bm q},z) = z - {\bm p}\cdot{\bm q}/m + i/\tau + \frac{(\alpha(z))^2}{z + {\bm p}\cdot{\bm q}/m + i/\tau}
\label{eq:16ad}
\ee
we find
\bea
{\tilde J}_0({\bm q},z) &=& \frac{i}{V} \sum_{\bm p} \delta f({\bm p},{\bm q},t=0)/N({\bm p},{\bm q},z)
\label{eq:16ab}\\
{\tilde J}({\bm q},z) &=& \frac{1}{4\pi} \int d\Omega_{\bm p}\ 1/N(\kF\hat{\bm p},{\bm q},z)
\label{eq:16ac}\\
K_0({\bm q},z) \!&=& \frac{i}{V} \sum_{\bm p} \frac{\delta f(-{\bm p},{\bm q},t=0)}{N({\bm p},{\bm q},z)(z + {\bm p}\cdot{\bm q}/m + i/\tau)}\ ,
\label{eq:16b}\\
K({\bm q},z) &=& \int \frac{d\Omega_{\bm p}}{4\pi} \frac{1}{N(\kF\hat{\bm p},{\bm q},z)(z + \vF\hat{\bm p}\cdot{\bm q} + i/\tau)}. \qquad\ 
\label{eq:16c}
\eea
\ese
Expanding the numerator in Eq.~(\ref{eq:16a}) to $O(q)$ and the denominator to $O(q^2)$, and neglecting corrections to the diffusion
pole that are analytic in $z$, we find 
\bse
\label{eqs:17}
\be
\delta n({\bm q}, z)  = \frac{i\delta n({\bm q},t=0)}{z + i D(z) q^2} + \frac{i\delta n_{\perp}({\bm q},t=0)}{[z + i D(z) q^2][1 + s_z \alpha(z)]}
\label{eq:17a}
\ee
with $\delta n_{\perp}$ from Eq.~(\ref{eq:3b}), and the renormalized diffusivity
\be
D(z) = D_0 s_z/[1 + s_z \alpha(z)\tau]\ .
\label{eq:17b}
\ee
\ese

The leading nonanalytic $z$-dependence 
of $\delta n$ follows from using Eq.~(\ref{eq:14d}) in Eqs.~(\ref{eqs:17}). 
In the asymptotic low-frequency regime $z \ll D_0 q^2$, the latter is $z^{1/2}$ in the first term on the right-hand side of Eq.~(\ref{eq:17a}), and
$z^{3/2}$ in the second one. For $z \gg D_0 q^2$, the first term has no nonanalytic $z$-dependence, while the second term goes as $z^{-1/2}$. 
The corresponding behavior in the time domain can be obtained by using the concept of generalized functions and their Fourier 
transforms \cite{Lighthill_1958, singularities_footnote}. 
All of the relevant nonanalyticities can be expressed in terms of  
\bse
\label{eqs:19}
\be
H_{\nu}(z) = -i\,[z^{\nu} - (-z)^{\nu}]\ .
\label{eq:19a}
\ee
The Laplace back transform for non-integer $\nu$ is \cite{Lighthill_1958}
\be
H_{\nu}(t) = \frac{-2}{\pi}\,\sin(\pi\nu) \sin(\pi\nu/2)\, \Gamma(1+\nu)/\vert t\vert^{1+\nu}\ ,
\label{eq:19b}
\ee
\ese
with $\Gamma$ the Gamma function. 
$H_{-1}(t)$ is a constant. Applying these results
to Eqs.~(\ref{eqs:17}) we obtain Eqs.~(\ref{eqs:4}).

We conclude with two general remarks. 

(1) A basic tenet of Statistical Mechanics is that a description that drastically reduces the number of degrees of freedom
is necessary, desirable, and physically sensible. In equilibrium one goes from a description in terms of $N = O(10^{23})$ variables 
to one in terms of two thermodynamic variables, say, density $n$ and energy density
$\epsilon$.  In a nonequlibrium fluid 
one also includes the fluid velocity ${\bm u}$. We have shown, using kinetic theory, that these descriptions are necessarily 
incomplete. One cannot simply integrate out the other
variables and still correctly describe transport or other nonequilibrium effects.

(2) There is no simple remedy for this incompleteness. A 
description in terms of the single-particle
distribution $f$, as used in this Letter, 
yields a solution in terms of the initial condition $f({\bm p},{\bm x},t=0)$.
This is not complete either, since in order to obtain a closed description in terms of $f$ alone one has to effectively
integrate out higher order distribution functions, such as 
the pair correlation function $G^{(2)} = f^{(2)} - f f$, with
$f^{(2)}$ the two-particle distribution function. A complete description in terms of $f$ therefore depends on $G^{(2)}(t=0)$, 
which in general is nonzero and also multiplies a LTT contribution \cite{Dorfman_vanBeijeren_Kirkpatrick_2021}. 
The conclusion is that, 
due to LTT effects, no reduced description can be truly complete.

To summarize: Long-time-tail effects have been investigated in a host of physical systems including classical fluids, superfluids, 
granular matter, active matter, electrons in solids, quark-gluon plasmas, and quantum gravity. The main message of this Letter 
is that a purely hydrodynamic description that is routinely used in all these fields is incomplete because non-hydrodynamic 
initial-condition effects cannot be ignored in general.

We thank Alex Levchenko for a discussion.


\begin{thebibliography}{37}
\expandafter\ifx\csname natexlab\endcsname\relax\def\natexlab#1{#1}\fi
\expandafter\ifx\csname bibnamefont\endcsname\relax
  \def\bibnamefont#1{#1}\fi
\expandafter\ifx\csname bibfnamefont\endcsname\relax
  \def\bibfnamefont#1{#1}\fi
\expandafter\ifx\csname citenamefont\endcsname\relax
  \def\citenamefont#1{#1}\fi
\expandafter\ifx\csname url\endcsname\relax
  \def\url#1{\texttt{#1}}\fi
\expandafter\ifx\csname urlprefix\endcsname\relax\def\urlprefix{URL }\fi
\providecommand{\bibinfo}[2]{#2}
\providecommand{\eprint}[2][]{\url{#2}}

\bibitem[{\citenamefont{Landau and Lifshitz}(1987)}]{Landau_Lifshitz_VI_1987}
\bibinfo{author}{\bibfnamefont{L.~D.} \bibnamefont{Landau}} \bibnamefont{and}
  \bibinfo{author}{\bibfnamefont{E.~M.} \bibnamefont{Lifshitz}},
  \emph{\bibinfo{title}{Fluid Mechanics}} (\bibinfo{publisher}{Pergamon,
  Oxford}, \bibinfo{year}{1987}).

\bibitem[{\citenamefont{Lifshitz and
  Pitaevskii}(1981)}]{Landau_Lifshitz_X_1981}
\bibinfo{author}{\bibfnamefont{E.~M.} \bibnamefont{Lifshitz}} \bibnamefont{and}
  \bibinfo{author}{\bibfnamefont{L.~P.} \bibnamefont{Pitaevskii}},
  \emph{\bibinfo{title}{Physical Kinetics}}
  (\bibinfo{publisher}{Butterworth-Heinemann, Oxford}, \bibinfo{year}{1981}).

\bibitem[{\citenamefont{Belmont et~al.}(2019)\citenamefont{Belmont, Rezeau,
  Riconda, and Zaslavsky}}]{Belmont_et_al_2019}
\bibinfo{author}{\bibfnamefont{G.}~\bibnamefont{Belmont}},
  \bibinfo{author}{\bibfnamefont{L.}~\bibnamefont{Rezeau}},
  \bibinfo{author}{\bibfnamefont{C.}~\bibnamefont{Riconda}}, \bibnamefont{and}
  \bibinfo{author}{\bibfnamefont{A.}~\bibnamefont{Zaslavsky}},
  \emph{\bibinfo{title}{Introduction to Plasma Physics}}
  (\bibinfo{publisher}{Elsevier}, \bibinfo{year}{2019}).

\bibitem[{\citenamefont{Forster}(1975)}]{Forster_1975}
\bibinfo{author}{\bibfnamefont{D.}~\bibnamefont{Forster}},
  \emph{\bibinfo{title}{Hydrodynamic Fluctuations, Broken Symmetry, and
  Correlation Functions}} (\bibinfo{publisher}{Benjamin, Reading, MA},
  \bibinfo{year}{1975}).

\bibitem[{\citenamefont{Martin et~al.}(1972)\citenamefont{Martin, Parodi, and
  Pershan}}]{Martin_Parodi_Pershan_1972}
\bibinfo{author}{\bibfnamefont{P.~C.} \bibnamefont{Martin}},
  \bibinfo{author}{\bibfnamefont{O.}~\bibnamefont{Parodi}}, \bibnamefont{and}
  \bibinfo{author}{\bibfnamefont{P.~S.} \bibnamefont{Pershan}},
  \bibinfo{journal}{Phys. Rev. A} \textbf{\bibinfo{volume}{6}},
  \bibinfo{pages}{2401} (\bibinfo{year}{1972}).

\bibitem[{\citenamefont{Marchetti et~al.}(2013)\citenamefont{Marchetti, Joanny,
  Ramaswamy, Liverpool, Prost, Rao, and Simha}}]{Marchetti_et_al_2013}
\bibinfo{author}{\bibfnamefont{M.~C.} \bibnamefont{Marchetti}},
  \bibinfo{author}{\bibfnamefont{J.~F.} \bibnamefont{Joanny}},
  \bibinfo{author}{\bibfnamefont{S.}~\bibnamefont{Ramaswamy}},
  \bibinfo{author}{\bibfnamefont{T.~B.} \bibnamefont{Liverpool}},
  \bibinfo{author}{\bibfnamefont{J.}~\bibnamefont{Prost}},
  \bibinfo{author}{\bibfnamefont{M.}~\bibnamefont{Rao}}, \bibnamefont{and}
  \bibinfo{author}{\bibfnamefont{R.~A.} \bibnamefont{Simha}},
  \bibinfo{journal}{Rev. Mod. Phys.} \textbf{\bibinfo{volume}{85}},
  \bibinfo{pages}{1143} (\bibinfo{year}{2013}).

\bibitem[{\citenamefont{Minwalla et~al.}(2011)\citenamefont{Minwalla, Hubeny,
  and Rangamani}}]{Minwalla_Hubeny_Rangamani_2011}
\bibinfo{author}{\bibfnamefont{S.}~\bibnamefont{Minwalla}},
  \bibinfo{author}{\bibfnamefont{V.~E.} \bibnamefont{Hubeny}},
  \bibnamefont{and}
  \bibinfo{author}{\bibfnamefont{M.}~\bibnamefont{Rangamani}}, in
  \emph{\bibinfo{booktitle}{Proceedings of the 2010 Theoretical Advanced Study
  Institute in Elementary Particle Physics}}, edited by
  \bibinfo{editor}{\bibfnamefont{M.}~\bibnamefont{Dine}},
  \bibinfo{editor}{\bibfnamefont{T.}~\bibnamefont{Banks}}, \bibnamefont{and}
  \bibinfo{editor}{\bibfnamefont{S.}~\bibnamefont{Sachdev}}
  (\bibinfo{publisher}{World Scientific, Singapore}, \bibinfo{year}{2011}), p.
  \bibinfo{pages}{817}.

\bibitem[{\citenamefont{Kovtun and Yaffe}(2003)}]{Kovtun_Yaffe_2003}
\bibinfo{author}{\bibfnamefont{P.}~\bibnamefont{Kovtun}} \bibnamefont{and}
  \bibinfo{author}{\bibfnamefont{L.~G.} \bibnamefont{Yaffe}},
  \bibinfo{journal}{Phys. Rev. D} \textbf{\bibinfo{volume}{68}},
  \bibinfo{pages}{025007} (\bibinfo{year}{2003}).

\bibitem[{\citenamefont{Mazeliauskas and
  Berges}(2019)}]{Mazeliauskas_Berges_2019}
\bibinfo{author}{\bibfnamefont{A.}~\bibnamefont{Mazeliauskas}}
  \bibnamefont{and} \bibinfo{author}{\bibfnamefont{J.}~\bibnamefont{Berges}},
  \bibinfo{journal}{Phys. Rev. Lett.} \textbf{\bibinfo{volume}{122}},
  \bibinfo{pages}{122301} (\bibinfo{year}{2019}).

\bibitem[{\citenamefont{Zaanen}(2016)}]{Zaanen_2016}
\bibinfo{author}{\bibfnamefont{J.}~\bibnamefont{Zaanen}},
  \bibinfo{journal}{Science} \textbf{\bibinfo{volume}{351}},
  \bibinfo{pages}{1026} (\bibinfo{year}{2016}).

\bibitem[{\citenamefont{Mehrer}(2007)}]{Mehrer_2007}
\bibinfo{author}{\bibfnamefont{H.}~\bibnamefont{Mehrer}},
  \emph{\bibinfo{title}{Diffusion in Solids}} (\bibinfo{publisher}{Springer,
  Berlin}, \bibinfo{year}{2007}).

\bibitem[{\citenamefont{Cussler}(1997)}]{Cussler_1997}
\bibinfo{author}{\bibfnamefont{E.~L.} \bibnamefont{Cussler}},
  \emph{\bibinfo{title}{Diffusion. Mass Transfer in Fluid Systems}}
  (\bibinfo{publisher}{Cambridge University Press, Cambridge, UK},
  \bibinfo{year}{1997}), \bibinfo{edition}{2nd} ed.

\bibitem[{\citenamefont{Stein}(1986)}]{Stein_1986}
\bibinfo{author}{\bibfnamefont{W.~D.} \bibnamefont{Stein}},
  \emph{\bibinfo{title}{Transport and Diffusion across Cell Membranes}}
  (\bibinfo{publisher}{Academic Press, Orlando, FL}, \bibinfo{year}{1986}).

\bibitem[{Lap()}]{Laplace_trafo_footnote}
\bibinfo{note}{Following Refs.~\onlinecite{Forster_1975} and
  \onlinecite{Zubarev_1960}, we define the Laplace transform $f(z)$ of a
  time-dependent function $f(t)$ by $f(z) = \pm \int_{-\infty}^{\infty}
  dt\,\Theta(\pm t) e^{izt}\,f(t)$, with $\pm$ for $\sgn\Im z \gtrless 0$.}

\bibitem[{\citenamefont{Dorfman et~al.}(2021)\citenamefont{Dorfman, van
  Beijeren, and Kirkpatrick}}]{Dorfman_vanBeijeren_Kirkpatrick_2021}
\bibinfo{author}{\bibfnamefont{J.~R.} \bibnamefont{Dorfman}},
  \bibinfo{author}{\bibfnamefont{H.}~\bibnamefont{van Beijeren}},
  \bibnamefont{and} \bibinfo{author}{\bibfnamefont{T.~R.}
  \bibnamefont{Kirkpatrick}}, \emph{\bibinfo{title}{Contemporary Kinetic Theory
  of Matter}} (\bibinfo{publisher}{Cambridge University Press},
  \bibinfo{year}{2021}).

\bibitem[{\citenamefont{Nieuwoudt et~al.}(1984)\citenamefont{Nieuwoudt,
  Kirkpatrick, and Dorfman}}]{Nieuwoudt_Kirkpatrick_Dorfman_1984}
\bibinfo{author}{\bibfnamefont{J.~C.} \bibnamefont{Nieuwoudt}},
  \bibinfo{author}{\bibfnamefont{T.~R.} \bibnamefont{Kirkpatrick}},
  \bibnamefont{and} \bibinfo{author}{\bibfnamefont{J.~R.}
  \bibnamefont{Dorfman}}, \bibinfo{journal}{J. Stat. Phys.}
  \textbf{\bibinfo{volume}{34}}, \bibinfo{pages}{203} (\bibinfo{year}{1984}).

\bibitem[{\citenamefont{Dorfman and Cohen}(1970)}]{Dorfman_Cohen_1970}
\bibinfo{author}{\bibfnamefont{J.~R.} \bibnamefont{Dorfman}} \bibnamefont{and}
  \bibinfo{author}{\bibfnamefont{E.~G.~D.} \bibnamefont{Cohen}},
  \bibinfo{journal}{Phys. Rev. Lett.} \textbf{\bibinfo{volume}{25}},
  \bibinfo{pages}{1257} (\bibinfo{year}{1970}).

\bibitem[{\citenamefont{Ernst et~al.}(1970)\citenamefont{Ernst, Hauge, and van
  Leeuwen}}]{Ernst_Hauge_van_Leeuwen_1970}
\bibinfo{author}{\bibfnamefont{M.~H.} \bibnamefont{Ernst}},
  \bibinfo{author}{\bibfnamefont{E.~H.} \bibnamefont{Hauge}}, \bibnamefont{and}
  \bibinfo{author}{\bibfnamefont{J.~M.~J.} \bibnamefont{van Leeuwen}},
  \bibinfo{journal}{Phys. Rev. Lett.} \textbf{\bibinfo{volume}{25}},
  \bibinfo{pages}{1254} (\bibinfo{year}{1970}).

\bibitem[{\citenamefont{Lee and Ramakrishnan}(1985)}]{Lee_Ramakrishnan_1985}
\bibinfo{author}{\bibfnamefont{P.~A.} \bibnamefont{Lee}} \bibnamefont{and}
  \bibinfo{author}{\bibfnamefont{T.~V.} \bibnamefont{Ramakrishnan}},
  \bibinfo{journal}{Rev. Mod. Phys.} \textbf{\bibinfo{volume}{57}},
  \bibinfo{pages}{287} (\bibinfo{year}{1985}).

\bibitem[{AA_()}]{AA_effects_footnote}
\bibinfo{note}{The same type of LTT arises by a different mechanism in
  interacting electron systems via an interplay between the interaction and
  quenched disorder, see Refs.~\onlinecite{Altshuler_Aronov_1979,
  Altshuler_Aronov_Lee_1980, Belitz_Kirkpatrick_1994}.}

\bibitem[{\citenamefont{Kohn and Luttinger}(1957)}]{Kohn_Luttinger_1957}
\bibinfo{author}{\bibfnamefont{W.}~\bibnamefont{Kohn}} \bibnamefont{and}
  \bibinfo{author}{\bibfnamefont{J.~M.} \bibnamefont{Luttinger}},
  \bibinfo{journal}{Phys. Rev.} \textbf{\bibinfo{volume}{108}},
  \bibinfo{pages}{590} (\bibinfo{year}{1957}).

\bibitem[{\citenamefont{Greenwood}(1958)}]{Greenwood_1958}
\bibinfo{author}{\bibfnamefont{D.~A.} \bibnamefont{Greenwood}},
  \bibinfo{journal}{Proc. Phys. Soc.} \textbf{\bibinfo{volume}{71}},
  \bibinfo{pages}{585} (\bibinfo{year}{1958}).

\bibitem[{\citenamefont{Abrahams et~al.}(1979)\citenamefont{Abrahams, Anderson,
  Licciardello, and Ramakrishnan}}]{Abrahams_et_al_1979}
\bibinfo{author}{\bibfnamefont{E.}~\bibnamefont{Abrahams}},
  \bibinfo{author}{\bibfnamefont{P.~W.} \bibnamefont{Anderson}},
  \bibinfo{author}{\bibfnamefont{D.~C.} \bibnamefont{Licciardello}},
  \bibnamefont{and} \bibinfo{author}{\bibfnamefont{T.~V.}
  \bibnamefont{Ramakrishnan}}, \bibinfo{journal}{Phys. Rev. Lett.}
  \textbf{\bibinfo{volume}{42}}, \bibinfo{pages}{673} (\bibinfo{year}{1979}).

\bibitem[{\citenamefont{Wegner}(1979)}]{Wegner_1979}
\bibinfo{author}{\bibfnamefont{F.}~\bibnamefont{Wegner}}, \bibinfo{journal}{Z.
  Phys. B} \textbf{\bibinfo{volume}{35}}, \bibinfo{pages}{207}
  (\bibinfo{year}{1979}).

\bibitem[{\citenamefont{Hershfield and
  Ambegaokar}(1986)}]{Hershfield_Ambegaokar_1986}
\bibinfo{author}{\bibfnamefont{S.}~\bibnamefont{Hershfield}} \bibnamefont{and}
  \bibinfo{author}{\bibfnamefont{V.}~\bibnamefont{Ambegaokar}},
  \bibinfo{journal}{Phys. Rev. B} \textbf{\bibinfo{volume}{34}},
  \bibinfo{pages}{2147} (\bibinfo{year}{1986}).

\bibitem[{\citenamefont{Strinati et~al.}(1989)\citenamefont{Strinati,
  Castellani, and Castro}}]{Strinati_Castellani_DiCastro_1989}
\bibinfo{author}{\bibfnamefont{G.}~\bibnamefont{Strinati}},
  \bibinfo{author}{\bibfnamefont{C.}~\bibnamefont{Castellani}},
  \bibnamefont{and} \bibinfo{author}{\bibfnamefont{C.~D.}
  \bibnamefont{Castro}}, \bibinfo{journal}{Phys. Rev. B}
  \textbf{\bibinfo{volume}{40}}, \bibinfo{pages}{12237} (\bibinfo{year}{1989}).

\bibitem[{\citenamefont{Gorkov et~al.}(1979)\citenamefont{Gorkov, Larkin, and
  Khmelnitskii}}]{Gorkov_Larkin_Khmelnitskii_1979}
\bibinfo{author}{\bibfnamefont{L.~P.} \bibnamefont{Gorkov}},
  \bibinfo{author}{\bibfnamefont{A.}~\bibnamefont{Larkin}}, \bibnamefont{and}
  \bibinfo{author}{\bibfnamefont{D.~E.} \bibnamefont{Khmelnitskii}},
  \bibinfo{journal}{Pis'ma Zh. Eksp. Teor. Fiz.} \textbf{\bibinfo{volume}{30}},
  \bibinfo{pages}{248} (\bibinfo{year}{1979}), \bibinfo{note}{[JETP Lett. {\bf
  30}, 228 (1979)]}.

\bibitem[{alp()}]{alpha_footnote}
\bibinfo{note}{This is an integral over the density relaxation in Boltzmann
  approximation. The integral extends over all wave numbers, but the integrand
  is diffusive only for small wave numbers. This observation accounts for the
  non-universal constant in Eq.~(\ref{eq:14d}) that depends on the density
  response in the non-hydrodynamic regime.}

\bibitem[{mod()}]{mode-coupling_footnote}
\bibinfo{note}{The object $M(z) = i/\tau + i \alpha(z)$ is identical with the
  memory kernel that appears in the mode-mode coupling theory of weak
  localization \cite{Vollhardt_Woelfle_1980, Belitz_Gold_Goetze_1981}.}

\bibitem[{\citenamefont{Lighthill}(1958)}]{Lighthill_1958}
\bibinfo{author}{\bibfnamefont{M.~J.} \bibnamefont{Lighthill}},
  \emph{\bibinfo{title}{Introduction to Fourier analysis and generalised
  functions}} (\bibinfo{publisher}{Cambridge University Press, Cambridge},
  \bibinfo{year}{1958}).

\bibitem[{sin()}]{singularities_footnote}
\bibinfo{note}{The algebraic decay at long times is entirely determined by the
  singularities at $z=0$ and the associated branch cuts \cite{Lighthill_1958}.
  In addition, there are poles in the complex-frequency plane that lead to
  exponential decay with a non-analytic wave-number dependence of the
  diffusivity.}

\bibitem[{\citenamefont{Zubarev}(1960)}]{Zubarev_1960}
\bibinfo{author}{\bibfnamefont{D.~N.} \bibnamefont{Zubarev}},
  \bibinfo{journal}{Usp. Fiz. Nauk} \textbf{\bibinfo{volume}{71}},
  \bibinfo{pages}{71} (\bibinfo{year}{1960}), \bibinfo{note}{[Sov. Phys. Usp.
  {\bf 3}, 320 (1960)]}.

\bibitem[{\citenamefont{Altshuler and Aronov}(1979)}]{Altshuler_Aronov_1979}
\bibinfo{author}{\bibfnamefont{B.~L.} \bibnamefont{Altshuler}}
  \bibnamefont{and} \bibinfo{author}{\bibfnamefont{A.~G.}
  \bibnamefont{Aronov}}, \bibinfo{journal}{Solid State Commun.}
  \textbf{\bibinfo{volume}{30}}, \bibinfo{pages}{115} (\bibinfo{year}{1979}).

\bibitem[{\citenamefont{Altshuler et~al.}(1980)\citenamefont{Altshuler, Aronov,
  and Lee}}]{Altshuler_Aronov_Lee_1980}
\bibinfo{author}{\bibfnamefont{B.~L.} \bibnamefont{Altshuler}},
  \bibinfo{author}{\bibfnamefont{A.~G.} \bibnamefont{Aronov}},
  \bibnamefont{and} \bibinfo{author}{\bibfnamefont{P.~A.} \bibnamefont{Lee}},
  \bibinfo{journal}{Phys. Rev. Lett.} \textbf{\bibinfo{volume}{44}},
  \bibinfo{pages}{1288} (\bibinfo{year}{1980}).

\bibitem[{\citenamefont{Belitz and
  Kirkpatrick}(1994)}]{Belitz_Kirkpatrick_1994}
\bibinfo{author}{\bibfnamefont{D.}~\bibnamefont{Belitz}} \bibnamefont{and}
  \bibinfo{author}{\bibfnamefont{T.~R.} \bibnamefont{Kirkpatrick}},
  \bibinfo{journal}{Rev. Mod. Phys.} \textbf{\bibinfo{volume}{66}},
  \bibinfo{pages}{261} (\bibinfo{year}{1994}).

\bibitem[{\citenamefont{Vollhardt and
  W{\"o}lfle}(1980)}]{Vollhardt_Woelfle_1980}
\bibinfo{author}{\bibfnamefont{D.}~\bibnamefont{Vollhardt}} \bibnamefont{and}
  \bibinfo{author}{\bibfnamefont{P.}~\bibnamefont{W{\"o}lfle}},
  \bibinfo{journal}{Phys. Rev. B} \textbf{\bibinfo{volume}{22}},
  \bibinfo{pages}{4666} (\bibinfo{year}{1980}).

\bibitem[{\citenamefont{Belitz et~al.}(1981)\citenamefont{Belitz, Gold, and
  G{\"o}tze}}]{Belitz_Gold_Goetze_1981}
\bibinfo{author}{\bibfnamefont{D.}~\bibnamefont{Belitz}},
  \bibinfo{author}{\bibfnamefont{A.}~\bibnamefont{Gold}}, \bibnamefont{and}
  \bibinfo{author}{\bibfnamefont{W.}~\bibnamefont{G{\"o}tze}},
  \bibinfo{journal}{Z. Phys. B} \textbf{\bibinfo{volume}{44}},
  \bibinfo{pages}{273} (\bibinfo{year}{1981}).

\end{thebibliography}

\end{document}